\documentstyle[12pt,epsfig]{article}
\begin{document}

May 1996  \hfill  DESY 96--095
\vspace{2cm}

\begin{center}

{\Large \bf{Possible Measurements of Single and}}   \\ [5pt]
{\Large \bf{Double Spin Asymmetries with {\it HERA--$\vec{N}$}
  \footnote{Updated version of a talk given at the "TRENDS IN
   COLLIDER SPIN PHYSICS" conference, ICTP Trieste, Dec. 4--8, 1995
   }}   \\ [40pt]

{\large Wolf-Dieter Nowak~\footnote{e-mail: nowakw@ifh.de}}    \\
[30pt]

{\small \it DESY-Institut f\"ur Hochenergiephysik \\  [5pt]
Platanenallee 6   \\
 D-15738 Zeuthen, Germany }             \\ [3cm]

}
\vspace{1cm}

{\bf Abstract}
\end{center}
\smallskip

  \footnotesize{
    The physics scope of a possible future experiment 
    utilizing an internal polarized nucleon target in the HERA proton
    beam is discussed.
    By measuring single spin asymmetries in inclusive particle production 
    at 820~GeV beam energy the higher-twist sector of 
    perturbative QCD can be probed with good statistical sensitivity.
    To support the physics case for proton polarization at HERA, we
    consider the measurement of double spin asymmetries
    in photon plus jet production. It appears possible to determine
    the polarized gluon distribution in the range 
    0.1~$\leq x_{gluon} \leq$~0.4 with good accuracy.}

\newpage

  \section{Introduction}

An experiment ('{\it HERA--$\vec{N}$}')
utilising an internal polarized nucleon target in the
820~GeV HERA proton beam would constitute a logical extension of the
study at DESY, commenced with the {\it HERMES} experiment,
of the nucleon structure by investigating in detail its spin degrees
of freedom.
Judged from today, this would be the only other place to study high
energy nucleon--nucleon spin physics besides the envisaged dedicated
RHIC spin program at BNL which is supposed to get started in about
five years from now.  \\
The internal polarized nucleon target offers unique features such as polarization 
above 80\%, no or small dilution, and a high density up to $10^{14}$ 
atoms/cm$^2$ \cite{ste1}. Moreover, only small systematic errors are
expected when proton and neutron results extracted from 
hydrogen and deuterium gas data, respectively, are compared.  \\
As long as the polarized target is operated in the unpolarized proton
beam, {\it HERA--$\vec{N}$} would be focused on
measurements of single spin asymmetries in different inclusive final states
('Phase~I' physics) \cite{now2,Proc,nur2}.
Once polarized protons should become available at HERA,
the same set-up would be readily available to measure 
various kinds of double spin asymmetries. 
These 'Phase~II' measurements would constitute an alternative 
--~fixed target~-- approach to similar physics as it will be
accessible to the collider experiments {\it STAR} and {\it PHENIX}
at the low end of the RHIC energy scale ($\sqrt{s}~\simeq~50$~GeV)
\cite{RSC}. \\

In the following the physics capabilities of {\it HERA--$\vec{N}$} will be 
illustrated by selecting two specific measurements. As an example for Phase~I
physics, inclusive pion production will be shown to serve as a
sensitive test of the
QCD spin sector in the transition region between the non--perturbative and 
the perturbative regime. Photon plus jet production,
chosen as an example for Phase~II physics, will be shown to be
suitable for measuring the polarized gluon distribution with a statistical
accuracy comparable to that in future measurements at RHIC.

A more extended description of the present
understanding of the prospects of polarized nucleon--nucleon
physics at HERA is given in a separate paper
\cite{IntRep} which describes the results
of a recent workshop at DESY--Zeuthen.

\vspace{0.5cm}

\section{Expected Size of Asymmetries}

In the {\it unpolarized} HERA proton beam the physically interesting
single spin asymmetry $A_N$ arises from inverting the direction of
the transversely oriented target polarization vector:

\begin{equation}
\label{A_N}
\qquad  A_N = \frac{d\sigma(pN^{\uparrow}) - d\sigma(pN^{\downarrow})}
{d\sigma(pN^{\uparrow}) +  d\sigma(pN^{\downarrow})}
\end{equation}

$A_N$ is supposed to be very small in pQCD \cite{siv1}, as
will be discussed below in more detail.

In a {\it polarized} proton beam the double spin asymmetry of primary
interest is the normalized cross section difference between parallel and 
anti--parallel alignment of the longitudinally oriented beam and target
polarization vectors:

\vspace{-1ex}
\begin{equation}
\label{A_LL}
\qquad  A_{LL} = \frac{d\sigma(\vec{p}\vec{N}) 
- d\sigma(\vec{p} \stackrel{\leftarrow}{N})}{d\sigma(\vec{p}\vec{N}) 
 + d\sigma(\vec{p}\stackrel{\leftarrow}{N})}
\end{equation}

To assess the order of magnitude of an asymmetry measureable in the
hadronic reaction $A + B \rightarrow C + (D) + X$
it has to be taken into account that on the parton level,
usually several subprocesses $a + b \rightarrow c + d$
contribute. Utilizing the QCD factorization property the measured
asymmetry can be represented as

\newpage

\vspace{-1ex}
\begin{equation}
\label{A^had}
A^{had} \quad \sim \quad (P_{B}) \cdot P_{T} \cdot \sum_{a,b} 
\quad (A^a) \cdot A^b \cdot A^{ab} \cdot \Big [ A^{c}  \Big ]
\cdot \Big [ A^{d}  \Big ]
\end{equation}
\vspace{-1ex}

where $P_{B}$ and $P_{T}$ are the average beam and target
polarisation, respectively. Throughout the paper $P_{B} = 0.6$ 
and $P_{T} = 0.8$ are assumed.
The average parton polarization in a fully polarized nucleon
is believed to be $A^a,A^b \simeq 0.25 ... 0.5$. The hard scattering asymmetry
$A^{ab}$, completely calculable in pQCD for double spin asymmetries,
ranges between 0.01 and 1. No reliable numbers are known yet for 
the fragmentation asymmetries $A^c,A^d$; fortunately this dilution does not
appear when photon, Drell--Yan and/or jet production is considered.

Taking into account that $P_{B}$ and
$A^{a}$ do not appear for unpolarized beams when
inserting the above numbers into eq.(\ref{A^had}) we find
that the measurable asymmetry values 
might range from 0.001 to about 0.25.
Hence precision measurements with statistical sensitivities
of 0.01 or better are required to study the spin dependence of 
nucleon--nucleon collisions. At the same time a careful assessment of
systematic errors is indispensable.

\vspace{0.5cm}

\section{Luminosity and Sensitivity}

The statistical sensitivity in measuring a spin asymmetry A is given 
by \cite{now1}

\begin{equation}
\label{Delta_A}
\delta A = \frac{1}{P_{B} \cdot P_{T}} \cdot \frac{1}{\sqrt{
    C \cdot {\cal{L}} \cdot T}} \cdot \frac{1}{\sqrt{\sigma}} \; ,
\end{equation}

C~$\simeq~50\%$ represents the anticipated combined trigger and 
reconstruction efficiency.
Using $\bar{I}_B = 80 \; \mbox{mA} = 0.5 \cdot 10^{18} \; s^{-1}$ as a
realistic figure for the 1996 average HERA proton beam current and a
realistic polarized target density of 
$n_T = 3 \cdot 10^{13}$ atoms/cm$^2$ \cite{ste1}
the 
projected luminosity becomes

\vspace{-1em}
\begin{equation}
\label{Lumi}
{\cal{L}} = n_T \cdot \bar{I}_B = 1.5 \cdot 10^{31} \; \mbox{cm}^{-2}
s^{-1} \; .
\end{equation}
\vspace{-1.5em} 

Assuming for the total running time an equivalent of 
$\quad T = 1.6 \cdot 10^7 \;s$
at 100 \% efficiency the integrated luminosity is obtained as

\vspace{-1em}
\begin{equation}
\label{intLum}
{\cal{L}} \cdot T = 240 \; pb^{-1}
\end{equation}
\vspace{-2em} 

The resulting sensitivities are given by

\vspace{.5ex}
\begin{center}
\begin{tabular}{lp{8cm}}
$\delta A_{ssa} = 0.10 / \sqrt{\sigma / [pb]}$ &  for single
spin asymmetries and \\
\vspace{5pt}
$\delta A_{dsa} = 0.17 / \sqrt{\sigma / [pb]}$ & for double
spin asymmetries.
\end{tabular}
\end{center}

These sensitivities do not include the spectrometer acceptance.
In a typical fixed target experiment the main reduction is imposed by the 
limited azimuthal coverage which decreases the acceptance by about
$1/\sqrt{1/3}~\simeq~2$. On the other hand, 
experience from UA6 at CERN indicates that after having gained
some practical running experience it might turn out
feasible to operate the polarized gas target at a 2 or 3 times 
higher density without seriously affecting the proton beam lifetime. 
In addition, proton currents much higher than the original HERA design
value of 160~mA will presumably be provided later in a possible HERA high
luminosity scenario. Thus eventually a net improvement in sensitivity
by a factor of 2 or better can be realistically expected.

Using 1995 realistic HERA conditions (including 33\% 
combined up--time for accele\-ra\-tor and experiment)
the integrated luminosity calculated in eq.(\ref{intLum}) would correspond 
to about 3 calendar years of machine operation with 6 months 
physics running per year. Taking account of the improvements 
expected to have been made by the time of
a possible {\it HERA--$\vec{N}$} experiment as noted above,
leads to a running time of about
1 to 2 real years required to collect 240 pb$^{-1}$.  \\

\newpage

\section{Kinematics}

In fig.~1 the interdependence between laboratory angle and
laboratory momentum for inclusive
pion production is shown as a function of $x_F$ and $p_T$.  \\

\begin{figure}[htb]
\vspace*{-3.5cm}
\begin{center}
\epsfig{file=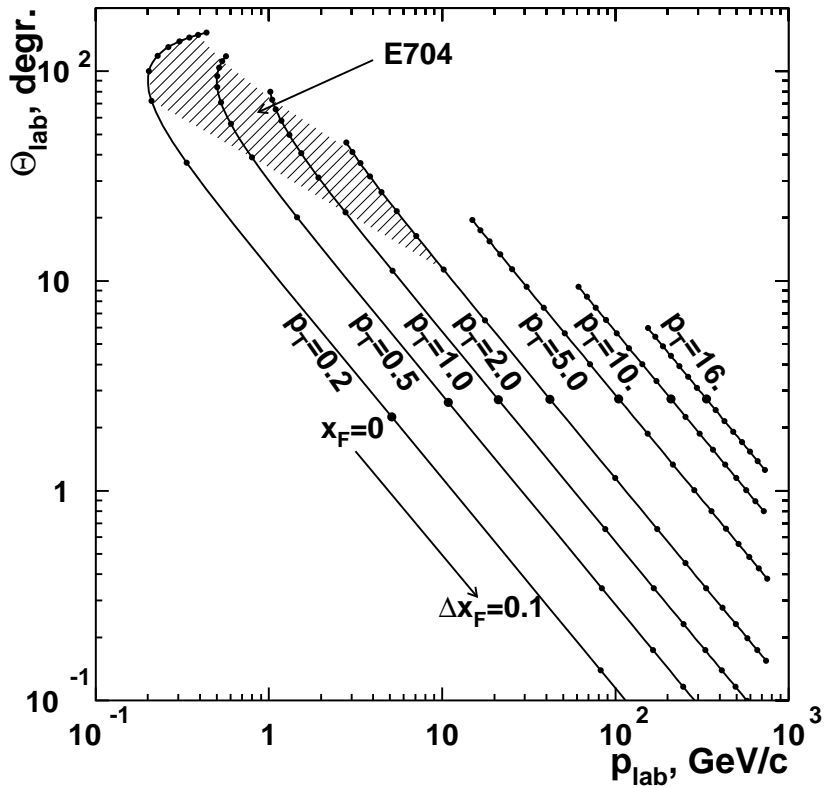,bbllx=60pt,bblly=100pt,bburx=480pt,bbury=650pt,
  width=8.5cm}
\end{center}

\vspace*{1.8cm}

\noindent
\caption[fig1]{\it Kinematics for inclusive pion production.}
\end{figure}
%
%

The hatched area corresponds to the fragmentation region of the
polarized nucleon, for which
the {\it E704} Collaboration has found significant non-zero values in the
pion single spin asymmetry \cite{704a}. For easier comparison this area
is drawn in the backward hemisphere here since 
during Phase~I of {\it HERA-$\vec{N}$} the {\it target} nucleon would carry the 
polarization, whereas in {\it E704} the polarized {\it beam} was hitting an 
unpolarized target. It is obvious that in order to check the {\it  E704} results 
and to measure the hitherto unknown $p_T$ dependence of single spin asymmetries
at large enough $|x_F|$ the spectrometer 
must be able to detect pions emitted 
under rather large laboratory angles ($\Theta_{lab} \geq 10$~degrees) 
which implies low momenta ($p_{lab} \leq$~10~GeV/c). 
We note that the measurement of singly polarized {\it elastic} scattering,
where significant non--zero results \cite{CERNela,AGSela} are also at
variance with the pQCD prediction $A_N \simeq 0$, requires recoil proton detection 
under large laboratory angles as well.  \\

In contrast, the study of double spin asymmetries during Phase~II of 
{\it HERA-$\vec{N}$} (e.g. in photon plus jet production)
would require to detect {\it both} photon and jet;
rather small laboratory angles (a few tens of milliradians) are involved.
Altogether it can be concluded that the {\it HERA--$\vec{N}$} apparatus
should have a wide-aperture spectrometer to ensure that a sufficiently broad
spin physics program can be realized.

\vspace{0.5cm}

\section{Single Spin Asymmetries}

Single spin asymmetries in large $p_T$ inclusive production, both
in proton-nucleon and lepton-nucleon interactions have recently
received much attention \cite{ssaList}. The naive expectation
from perturbative QCD
that they should be zero has been proven to be false,
both experimentally and theoretically.
It is now clear that twist-3 effects are responsible for
these asymmetries, which come out to be be zero only in leading twist-2
perturbative QCD.

Several models and theoretical analyses suggest higher-twist 
effects: there might be twist-3 dynamical contributions
(hard scattering higher-twists \cite{ter3});
there might also be intrinsic $k_\perp$ effects, both in the
quark fragmentation process \cite{col,art} and in the quark
distribution functions \cite{siv1,siv2,ans,ans1}. The latter are not by
themselves higher-twist contributions but rather non-perturbative
universal nucleon properties giving rise to twist-3 contributions
when convoluted with the hard scattering cross sections.
The dynamical contributions result from a short distance part
calculable in perturbative QCD 
combined with a long distance part related to quark-gluon
correlations \cite{ter3}.  \\

We consider the hadron level reaction $p + N^{\uparrow} \to h + X$
and a corresponding singly polarized partonic subprocess
$a + b^{\uparrow} \to c + d$. According to the leading twist QCD
factorization theorem the single spin asymmetry $A_N$ defined in
eq. (\ref{A_N}) can be represented as

\begin{equation}
\label{A_Nssa}
A_{N}  \sim \sum_{ab \rightarrow cd} f_{a/p} \otimes \Delta_{T}f_{b/N}
\otimes \hat{a}_{N} \otimes D_{h/c}
\end{equation}

with $\Delta_T f_{b/N} =
f_{b^{\uparrow}/N^{\uparrow}}-f_{b^{\downarrow}/N^{\uparrow}}$.
The number densities of 'beam partons' $a$ and polarized 
'target partons' $b^{\uparrow}$ in their parent nucleon
are denoted by $f_{a/p}$ and $f_{b^\uparrow/N^{\uparrow}}$, respectively.
$D_{h/c}$ is the number density of the hadron $h$ from
fragmentation of the outgoing parton $c$. 
Finally, $\hat a_N$ is the single spin asymmetry for the parton
subprocess under consideration, which is zero, thus giving $A_N = 0$.

However, as we said above, intrinsic $k_\perp$ and higher-twist
effects might modify the above equation leading to twist-3 
non-zero contributions to $A_N$ ranging between 10 and 20\%.

It is tempting to try a separation of these possible contributions to
single spin asymmetries by considering together the results of different
reactions and final states, respectively:

\vskip 12pt
$\bullet \quad pN^\uparrow \to hX$
\vskip 6pt

In this process, measurable with {\it HERA--$\vec{N}$},
all kinds of higher-twist contributions may be present;
those related to $k_\perp$ effects in the fragmentation function
have been considered in Ref. \cite{art} and those related to $k_\perp$
effects in the distribution function in Refs. \cite{ans,ans1}.
This asymmetry alone could not help in evaluating the relative
importance
of the different terms.

\vskip 12pt
$\bullet \quad pN^\uparrow \to \gamma X \quad$ or $\quad pN^\uparrow \to
\mu^+\mu^- X$
\vskip 6pt

These reactions, measurable with {\it HERA--$\vec{N}$} as well,
contain no fragmentation process. Possible
sources of non-zero single spin asymmetries are in the hard scattering
process or the quark distribution functions.

\vskip 12pt
$\bullet \quad lN^\uparrow \to hX$
\vskip 6pt

In such a process, measurable with {\it HERMES} \cite{her1} or 
{\it COMPASS} \cite{comp1}, the single spin asymmetry can originate 
either from
hard scattering or from $k_\perp$ effects in the fragmentation function,
but not in the distribution functions, as soft initial state interactions
are suppressed by powers of $\alpha_{em}$.

\vskip 12pt
$\bullet \quad lp^\uparrow \to \mu^+\mu^- X, \qquad
lp^\uparrow \to \gamma X, \qquad \gamma p^\uparrow \to \gamma X$
\vskip 6pt

Each of these processes yields a single spin asymmetry which cannot
originate from distribution or fragmentation $k_{\perp}$ effects;
it may only be due to higher-twist hard scattering effects, which
would thus be isolated. The first two processes can 
--~at least in principle~-- be measured with {\it HERMES} or {\it COMPASS},
the last one possibly using quasi-real photons with {\it H1} or {\it ZEUS}.
It is not obvious, however, that the counting rate for these processes
will be sufficient to utilize its apparent 'theoretical
usefulness' for disentangling the different fundamental
components involved in single spin asymmetries.

It is clear from the above discussion that a careful and complete
study of single spin asymmetries in several processes might be a
unique way of understanding the origin and importance of twist-3
contributions in inclusive hadronic interactions. In addition
it might also allow a determination of fundamental non-perturbative
properties of quarks inside polarized nucleons and of polarized
quark fragmentation. Such properties should be of universal value
and applicability and their knowledge might be as important as the
knowledge of unpolarized distribution and fragmentation functions.  \\

\begin{figure}[htb]
\vspace*{-9cm}
\begin{center}
\epsfig{file=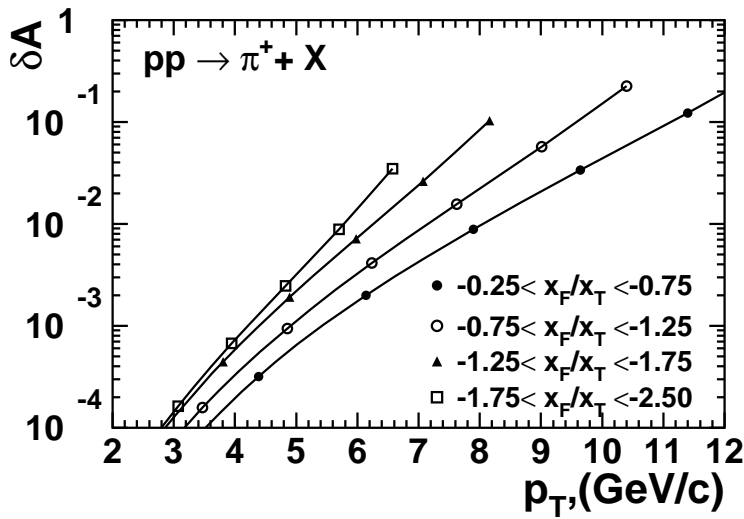,bbllx=60pt,bblly=100pt,bburx=480pt,bbury=650pt,
  width=10cm}
\end{center}

\vspace*{2.5cm}

\noindent
\caption[fig2]{\it Projected statistical accuracy vs. $p_T$ in different
       sectors of $\tilde{x}_F = x_F / x_T$.}
\end{figure}

We calculated the statistical sensitivity for inclusive pion
production under the above described {\it HERA--$\vec{N}$} conditions
\cite{IntRep}. Studying the (Monte Carlo) data in a 2-dimensional
$(p_T, x_F)$ binning with a cell size of 
$\Delta p_T \times \Delta p_L = 2 \times 2$ GeV$^2$ it turns out that
the $p_T$ values accessible to {\it HERA--$\vec{N}$} are significantly larger
than in the {\it E704} experiment.
The combined $p_T$ dependence of all higher-twist effects involved
can be measured with good accuracy ($\delta A_N \leq 0.05$)
up to transverse momenta of about 10~GeV/c in the central region $|x_F| < 0.2$
and up to 6~GeV/c in the target fragmentation region.

To make sure that the genuine $p_T$ dependence of the single spin asymmetry
is not spoiled by other implicit dependencies we write it in its most general
form as predicted by kinematics and dimensional analysis

\vspace{-2ex}
\begin{eqnarray}
\label{at}
A_N={M\over p_T} \cdot f(x_F,x_T)
\end{eqnarray}
\vspace{-1em}

with $x_T = p_T / p_T^{max} = 2 p_T / \sqrt{s}$. 
In a good approximation the energy dependence of $x_T$
can be neglected allowing for the simplification

\vspace{-2ex}
\begin{eqnarray}
\label{at1}
A_N={M\over p_T} \cdot f(\tilde x_F)
\end{eqnarray}
\vspace{-1em}

where $\tilde{x}_F = x_F / x_T$.
This behaviour can be clearly checked if the data is binned in an appropriate
way, i.e. within angular sectors having about the same $\tilde{x}_F$ value.
When considering a certain region of
high, i.e. target fragmentation like $x_F$ values (say, -0.4 to -0.8),
each angular sector covers
a rather small range in $p_T$, only. Each of these sectors being characterized
by its average $\tilde{x}_F$ then delivers its own $p_T$ dependence which
is scaled among each other by the difference in the function $f(\tilde{x}_F)$.
Considering all $\tilde{x}_F$ bins together, 
a clear picture is supposed to emerge on the
overall $p_T$ dependence of higher-twist effects. The sensitivity within each
sector vs. $p_T$ is displayed in fig.~2. As can be seen, there is good
sensitivity ($\delta A_N \leq 0.05$) up to $p_T \simeq 10$~GeV, hence 
the kinematical range accessible to {\it HERA--$\vec{N}$} is well
suited to 'scan' over the onset region of perturbative QCD.

\vspace{0.5cm}

\section{Double Spin Asymmetries}

Perturbative QCD allows for a simple calculation of Born double spin
asymmetries for various 2$\rightarrow$2 subprocesses at the partonic
level. The one-loop radiative corrections to various subprocesses have
now been calculated \cite{cont}, they produce only small changes in
the asymmetry in comparison with the leading order. Relying on
factorization a rich spectrum of hadronic level asymmetries is predicted
which constitutes the backbone of the RHIC spin physics program \cite{aki1}.

The measurement of $A_{LL}$ in certain final states (e.g. photon plus jet) 
is presently considered to be one of the most promising methods to
determine the (normalized) polarized gluon distribution $\Delta G / G$. 
Hence in the following
an estimate is given for the {\it HERA--$\vec{N}$} sensitivity to perform
such a measurement in the doubly polarized mode ('Phase~II').
When both photon and jet are detected in the hadronic final state the
Gluon--Compton subprocess $q~+~g~\rightarrow~\gamma~+~q$ dominates over the
quark--antiquark annihilation $q~+~\bar q~\rightarrow~\gamma~+~g$ if
the transverse momentum exceeds a few GeV/c. Describing the former one
by partonic formula \cite{IntRep} yields for the polarized and
unpolarized cross sections

\vspace{-2ex}
\begin{eqnarray}
\label{exclsim}
\Delta \sigma \quad = \quad \sum_f e^2_f \cdot \Delta q \cdot \Delta G
\cdot \Delta \hat \sigma \quad \sim \quad g_1 \cdot \Delta G \cdot 
\Delta \hat \sigma,\\
\sigma \quad =\quad \sum_f e^2_f \cdot q \cdot G \cdot
\hat \sigma \quad \sim \quad F_1 \cdot G \cdot \hat \sigma,
\end{eqnarray}
\vspace{-1em}

respectively. Hence the double spin asymmetry eq.(\ref{A_LL}) can be written as

\vspace{-1.5em}
\begin{eqnarray}
\label{asy}
A_{LL}=A_{DIS} \cdot \hat a_{LL} \cdot {{\Delta G}\over {G}}
\end{eqnarray}
\vspace{-1.5em}

Here $A_{DIS} = g_1 / F_1$ is the asymmetry measured in doubly
polarized lepton nucleon scattering and
the partonic level asymmetry $\hat a_{LL}$
is completely calculable in pQCD.
The absolute statistical error of $\Delta G / G$ 
is readily obtained as

\vspace{-1em}
\begin{eqnarray}
\label{err}
\delta [{{\Delta G} \over {G}} ]=
{{\delta A_{LL}}\over {A_{DIS} \cdot \hat a_{LL}}}.
\end{eqnarray}

\newpage

\begin{figure}[htb]
\vspace*{-3.5cm}
\begin{center}
\epsfig{file=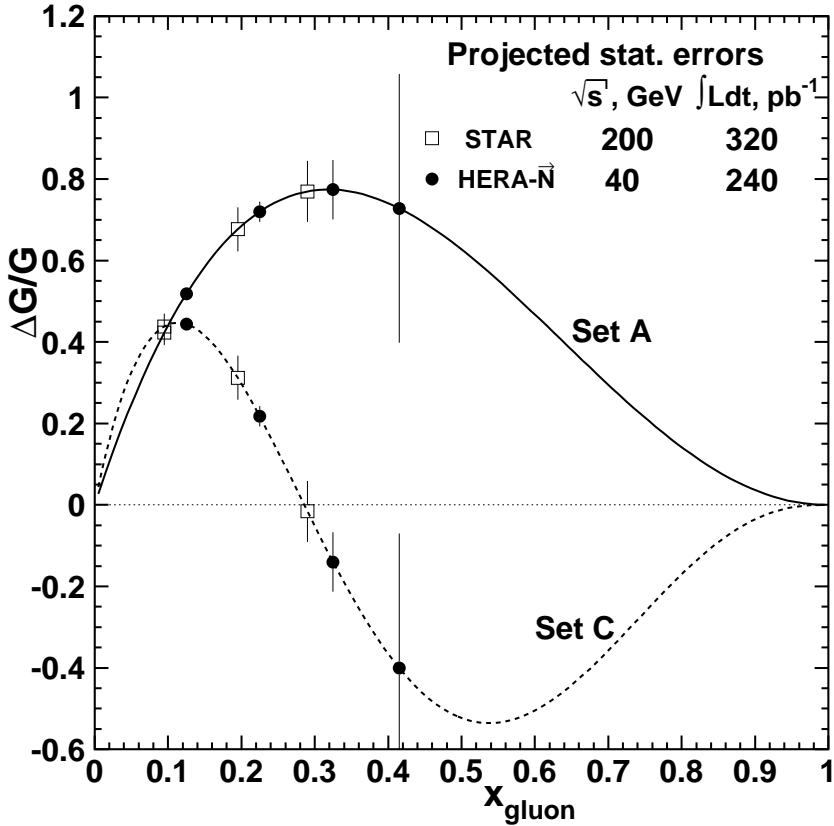,bbllx=60pt,bblly=100pt,bburx=480pt,bbury=650pt,
  width=9cm}
\end{center}

\vspace*{1.6cm}
\caption [fig3]{\it Typical predictions for the polarized gluon
  distribution confronted with the projected statistical errors expected for
       the {\it HERA-$\vec{N}$} and STAR experiments.}
\end{figure}

In fig.~3 we reproduce two typical predictions for the polarized gluon
distribution (Gehrmann and Stirling \cite{GS}, set A and C)
in conjunction with the projected {\it HERA--$\vec{N}$} statistical errors.
The errors show clearly that in the region
$0.1 \leq x_{g} \leq 0.4$ a significant result can be expected.
This statement will very probably remain valid if once the
systematic errors will have been estimated. Note that in this first
approximation the jet and photon reconstruction efficiences were not included. 
From preliminary jet studies \cite{IntRep} it can be anticipated that
a serious deterioration will occur only for lower $p_T$-values,
i.e. for the left-most point in fig.~3.
For comparison, we show the corresponding projected statistical errors for
{\it STAR} running at 200 GeV c.m. energy~\cite{yok3}. As can be seen, the
measurement of $\Delta G / G$ in doubly polarized nucleon-nucleon
collisions at HERA will yield results
competitive with those predicted for RHIC. As a word of caution we note
that the projected {\it STAR} errors available today 
are based upon much more detailed
studies than those shown here for {\it HERA--$\vec{N}$}.

\section*{Acknowledgements}

I am indebted to M. Anselmino and O. Teryaev for many
enlightening discussions which largely influenced the shape of
the above sketched {\it HERA--$\vec{N}$} physics program.
Many thanks are due to A. Jgoun, V. Korotkov, O. Martin and A. Tkabladze
whose contributions were essential to estimate
the statistical sensitivity of a possible future
polarized nucleon--nucleon experiment at HERA. Special thanks to
V. Korotkov for substantial help in preparing the figures. I thank
S. Nurushev for clarifying the details of the {\it E704}
measurements. The continous support of P. S\"oding
and his comments on the paper are warmly acknowledged.

\newpage



\begin{thebibliography}{9}


\bibitem{ste1}
E. Steffens, K. Zapfe--D\"uren, Proceedings of the 
{\it Workshop on the Prospects of Spin Physics at HERA},
Zeuthen, August 28-31, 1995, DESY 95--200, ed. by J.Bl\"umlein and
W.--D. Nowak, p.57.

\bibitem{now2}
W.--D. Nowak, ibid., p.145


\bibitem{Proc}
Proceedings of the 2nd Meeting on {\it Possible Measurements of Singly
Polarized $p\vec{p}$ and $p\vec{n}$ Collisions at HERA}, Zeuthen,
Aug.31--Sept.2, 1995, Desy Zeuthen Internal Report 95-05, ed. by
H. B\"ottcher and W.--D. Nowak

\bibitem{nur2}
S. B. Nurushev, ibid., p. 3

\bibitem{RSC}
RHIC Spin Coll., {\it Proposal on Spin Physics using the RHIC
Polarized Collider}, August 1992

\bibitem{IntRep}
M. Anselmino et al., Internal Report DESY--Zeuthen 96--04, May 1996

\bibitem{siv1}
D. Sivers, AIP Conference Proceedings 51, Particles and Fields
Subseries,
No. 17, {\it High Energy Physics with Polarized Beams and Polarized
  Targets},
Argonne, 1978, ed. by G. H. Thomas, p.505  \\
D. Sivers, {\it Phys. Rev. D 41}, 83 (1990)

\bibitem{now1}
W.-D. Nowak, AIP Conference Proceedings 343,
{\it High Energy Spin Physics, 11th International Symposium},
Bloomington, Indiana, Sept. 1994, ed. by K. Heller and S. Smith, p. 412

\bibitem{704a}
D. L. Adams et al., {\it Phys. Lett. B 264}, 462 (1991) \\
D. L. Adams et al., {\it Z. Phys. C56}, 181 (1992)

\bibitem{CERNela}
J. Antille et al., {\it Nucl. Phys. B185}, 1 (1981)

\bibitem{AGSela}
D. G. Crabb et al., {\it Nucl. Phys. B121}, 231 (1977) \\
P. H. Hansen et al., {\it Phys. Rev. Lett. 50}, 802 (1983) \\
D. C. Peaslee et al., {\it Phys. Rev. Lett. 51}, 2359 (1983) \\
P. R. Cameron et al., {\it Phys. Rev. D 32}, 3070 (1985) \\
D. G. Crabb et al., {\it Phys. Rev. Lett. 65}, 3241 (1990)

\bibitem{ssaList}
see references in \cite{IntRep}

\bibitem{ter3}
O. V. Teryaev, Proceedings of the 
{\it Workshop on the Prospects of Spin Physics at HERA},
Zeuthen, August 28-31, 1995, DESY 95--200, ed. by J.Bl\"umlein and
W.--D. Nowak, p.132

\bibitem{col}
J. Collins, {\it Nucl. Phys. B 396}, 161 (1993)

\bibitem{art}
X. Artru, J. Czyzewski and H. Yabuki, preprint LYCEN/9423 and TPJU
12/94,
May 1994, hep-ph/9405426

\bibitem{siv2}
D. Sivers, {\it Phys. Rev. D 41}, 261 (1991)

\bibitem{ans}
M. Anselmino, M. Boglione and F. Murgia, AIP Conference Proceedings 343,
{\it High Energy Spin Physics, 11th International Symposium},
Bloomington, Indiana, Sept. 1994, ed. by K. Heller and S. Smith, p. 446

\bibitem{ans1}
M. Anselmino, M. Boglione, F. Murgia,
{\it Phys. Lett. B 362}, 164 (1995)

\bibitem{her1}
HERMES Coll., P. Green et al., HERMES Technical Design Report,
{\it DESY-PRC 93/06}, July 1993

\bibitem{comp1}
COMPASS Coll., G. Baum et al., COMPASS Proposal,
{\it CERN/SPSLC/96-14}, SPSLC P297, 1 March 1997

\bibitem{cont}
A.P. Contogouris, B. Kamal, O. Korakianitis, F. Lebessis,
Z. Merebashvili,
{\it Nucl. Phys. Proc. Suppl. 39 BC}, 98 (1995)

\bibitem{aki1}
A. Yokosawa, AIP Conference Proceedings 343,
{\it High Energy Spin Physics, 11th International Symposium},
Bloomington, Indiana, Sept. 1994, ed. by K. Heller and S. Smith,
p. 841

\bibitem{GS}
T. Gehrmann and W. J. Stirling, {\it Durham University preprint},
DTP/95/82, (1995).

\bibitem{yok3}
A. Yokosawa, {\it these proceedings}

  \end{thebibliography}
\end{document}